\documentclass[5p]{elsarticle}
\usepackage{hyperref}
\usepackage{amsmath,bm}
\usepackage[utf8]{inputenc}
\usepackage[toc,page]{appendix}
\usepackage{graphicx}
\usepackage{float}
\usepackage{graphicx}
\usepackage{threeparttable}
\usepackage[abbreviations]{glossaries-extra}
\usepackage{epstopdf}
\usepackage{url}
\usepackage{subcaption}
\usepackage{framed} 
\usepackage{pgfplots}
\pgfplotsset{compat=1.14}
\usepackage{dcolumn}
\usepackage{bm}
\usepackage{subcaption}

\journal{...}

\begin{document}
\begin{frontmatter}
\title{Modeling ice crystal growth using the lattice Boltzmann method}
\author[OvGU]{Q.~Tan}
\author[OvGU,ETHZ]{S.A.~Hosseini\corref{mycorrespondingauthor}}
\cortext[mycorrespondingauthor]{Corresponding author}
\ead{seyed.hosseini@ovgu.de}
\author[MPI]{A.~Seidel-Morgenstern}
\author[OvGU]{D.~Th\'evenin}
\author[MPI]{H.~Lorenz}

\address[OvGU]{Laboratory of Fluid Dynamics and Technical Flows, University of Magdeburg ``Otto von Guericke'', D-39106 Magdeburg, Germany}
\address[ETHZ]{Department of Mechanical and Process Engineering, ETH Z\"urich, 8092 Z\"urich, Switzerland}
\address[MPI]{Max Planck Institute for Dynamics of Complex Technical Systems (MPI DCTS), 39106 Magdeburg, Germany}

\begin{abstract}
Given the multitude of growth habits, pronounced sensitivity to ambient conditions and wide range of scales involved, snowflake crystals are one of the most challenging systems to model. The present work focuses on the development and validation of a coupled flow/species/phase solver based on the lattice Boltzmann method. It is first shown that the model is able to correctly capture species and phase growth coupling. Furthermore, through a study of crystal growth subject to ventilation effects, it is shown that the model correctly captures hydrodynamics-induced asymmetrical growth. The validated solver is then used to model snowflake growth under different ambient conditions with respect to humidity and temperature in the plate-growth regime section of the Nakaya diagram. The resulting crystal habits are compared to both numerical  and experimental reference data available in the literature. The overall agreement with experimental data shows that the proposed algorithm correctly captures both the crystal shape and the onset of primary and secondary branching instabilities. As a final part of the study the effects of forced convection on snowflake growth are studied. It is shown, in agreement with observations in the literature, that under such condition the crystal exhibits non-symmetrical growth. The non-uniform humidity around the crystal due to forced convection can even result in the coexistence of different growth modes on different sides of the same crystal.
\end{abstract}

\end{frontmatter}

\section{Introduction}
\begin{figure}[H]                           
	\begin{subfigure}{0.3\textwidth}
	\includegraphics[width=1.5\textwidth]{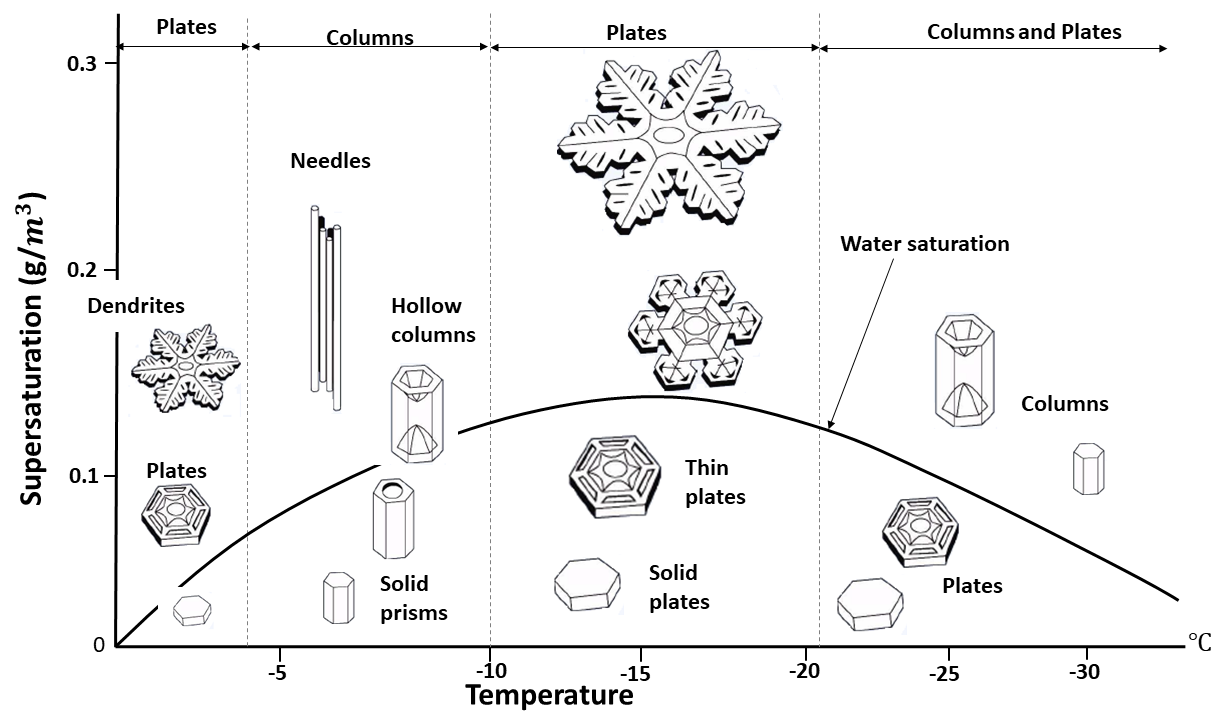}
	\end{subfigure}
	\caption{The snow crystal morphology diagram, showing the morphology of ice crystals growing from water vapor in air at 1 bar as a function of temperature and supersaturation. This figure is taken from~\cite{RN16,nakaya1954snow}.}
	\label{1}
\end{figure}

Ice crystals and their growth are of interest in many fields such as environmental sciences, agriculture and industries as aviation (encountering de-icing issues). Due to the wide spectrum of habits they take on, they have been the topic of scientific research for decades~\cite{frank1982snow,mason1992snow,wang2002shape}. Johannes Kepler was the first person to explore the growth mechanisms of snowflakes in 1611, and attempted to explain the possible origins of snow crystal symmetry~\cite{shafranovskii197514}. With the development of photography in the late 19th century, Wilson Bentley~\cite{chickering1864cloud} collected in 1931 thousands of snow crystal images~\cite{bentley1931snow}. In 1938, Ukichiro Nakaya and his co-workers~\cite{nakaya2013snow,gonda1997dendritic,nakaya1951formation} conducted comprehensive experimental studies of ice crystal growth to determine the relationship between growth conditions and crystal shape. The different growth modes and associated habits as a function of temperature and supersaturation (humidity) are illustrated in Fig.~\ref{1}.
Overall, based on the dominant growth direction, snow-flakes can be classified as pertaining to one of these two categories: plates, or columns/needles. The growth mode is essentially dictated by the temperature, while growth rate, and the associated instabilities, are affected by both temperature and humidity. As observed in Fig.~\ref{1}, crystal habits alternate between flat and columnar. Transitions between the different modes, {i.e.} plate and column, occur at around -5, -10 and -20$^\circ$C. Furthermore, higher humidity and thus, higher supersaturation as crystallization driving force contributes to more pronounced instability effects and therefore more complex shapes. Despite the efforts by Ukichiro Nakaya~\cite{nakaya1951formation} and other researchers, much of the phenomenology behind the growth of snowflakes still remains uncertain~\cite{libbrecht2005physics,libbrecht2001morphogenesis,libbrecht2017physical,ross2004snowflake}. In 1958, Mason~\cite{mason1958growth} suggested that the basic habit is determined by the surface diffusion rates. Later, Nelson and Knight~\cite{nelson1998snow} suggested that layer nucleation rates influence the morphology of snow crystals. Libbrecht~\cite{libbrecht2017physical} suggested that the different crystal habits and instabilities appearing during growth are direct consequences of two main competing mechanisms: vapor diffusion in air and kinetics of the water molecules attachment to the crystal surface~\cite{gonda1997dendritic,nelson2001growth,libbrecht2013measurements}. The geometrically ordered potential field on the crystal surface dictates the growth into faceted structures (resulting from the molecules' arrangement within the crystal lattice), while diffusion contributes to growth instabilities that produce dendritic branching and associated complex structures. Attempts to further explain the growth mechanism of snowflakes are still ongoing.
\par While most of the previous research was focused on experimental studies of ice growth, widespread effort has also been put on developing mathematical models and numerical tools. Microscopic models and simulations such as Molecular Dynamic simulations~\cite{magono1966meteorological} have been conducted to analyze fundamental processes such as surface adsorption and diffusion. While physically sound, these simulations are limited in time and space and are not efficient for simulations of a full snowflake (especially at later stages when instabilities appear and the crystal grow in size). So-called mesoscopic formulations are an interesting alternative. The cellular automaton of Gravner and Griffeath ~\cite{gravner2009modeling} is an illustration of such approaches. While providing spectacular results in predicting faceted growth, such models lack established connections with physical processes and parameters~\cite{kelly2014physical}. Furthermore, reliable models introducing additional physics such as interaction with a flow field are yet to be developed~\cite{chatterjee2006hybrid}. At the macroscopic level, crystal growth is modeled using either sharp or diffuse interface formulations. Various snowflake morphologies were simulated by Barrett et al.~\cite{barrett2012numerical} with a sharp interface model. However, only small supersaturations could be considered because of the numerical cost of the interface parametrization~\cite{singer2008phase}. In recent years, the phase-field  model~\cite{karma1998quantitative}, a type of diffuse interface formulation, has become one of the most popular methods for the simulation of crystal growth. The phase-field model is a powerful tool to simulate interface development in the crystallization process as the model does not require explicit tracking of the interface via front-tracking algorithms. In addition, the non-linear partial differential equations are obtained from the principles of non-equilibrium thermodynamics making the interface dynamics consistent without the need for explicit boundary treatments. Over the past several decades, few successful attempts are reported to model faceted snowflake growth~\cite{demange2017phase,demange2017growth}. Comparisons with experimental data were rather promising.
\newabbreviation{lbm}{LBM}{lattice Boltzmann method}
\newabbreviation{lb}{LB}{lattice Boltzmann}
\newabbreviation{ns}{NS}{Navier-Stokes}
\newabbreviation{pde}{PDE}{partial differential equation}
\par The \gls{lbm}, developed over the past decades, has become a popular alternative to classical solvers for the \gls{ns} equations ~\cite{doolen1998lattice}. It has since been extended to a variety of applications and flow regimes such as multi-phase flows~\cite{zheng2006lattice}, flows in porous media, turbulence and multi-component flows~\cite{hosseini2018mass,hosseini2019hybrid,hosseini2020low}. It has also been used, in combination with classical solvers for the solid phase, to simulate crystal growth~\cite{chatterjee2006hybrid,medvedev2005lattice,rasin2005phase,miller2006lattice,medvedev2006influence,huber2008lattice,sun2009lattice,lin2014three}. While initially developed as a discrete solver for the Boltzmann equation in the hydrodynamic regime, it has also widely been used as a solver for different types of parabolic \gls{pde} via appropriate parametrization of the equilibrium state and the collision-streaming operators. In that same spirit a number of work have proposed \gls{lbm}-based formulations to solve the phase-field evolution equation.  Younsi et al.~\cite{younsi2016anisotropy} modified the standard \gls{lb} equations and  equilibrium distribution function to introduce anisotropic surface tension and growth rate effects. The proposed model was successfully used to simulate anisotropic and dendritic growth. While readily applied to generic systems, these models have rarely been used to simulate realistic systems such as snowflake growth.
\par The aim of the present work is to present a pure \gls{lb} model able to correctly capture not only different growth modes of snowflakes in the platelet regime, but also effects caused by the surrounding convective field. For this purpose, after an overview of the approach proposed to model crystal growth, it will first be validated against generic test-cases well-documented in the literature (both with and without ventilation effects). The solver is then used to model the growth of a single ice crystal under different conditions regarding temperature and supersaturation. The obtained shapes are validated (qualitatively) against both experimental and numerical data available in the literature. It will also be shown that the growth modes are in agreement with those predicted in the Nakaya diagram. Finally, the effect of fluid flow on dendritic growth is studied. Interestingly, rarely studied growth habits are observed in the simulations;For example triangular shapes as observed only when crystals are exposed to a background convection.
\section{Theoretical background}
While influenced by both temperature and humidity (referred to as supersaturation in what follows), given the more pronounced effect of the latter on faceted growth and dendritic instabilities, the temperature field will be considered to be uniform during the present study. As such, the phase-field growth dynamics will involve two coupled equations (apart from the flow field solver). One will describe the morphology of the snowflakes through a phase parameter $\phi$; the second one is an advection/diffusion-type equation for the supersaturation field $U$.
\subsection{Diffuse-interface formulation: governing equations}
The kinetics of phase-field are described by a parameter $\phi$ designating solid and fluid phases respectively when reaching +1 (ice) and -1 (vapor) and the reduced supersaturation of water vapor defined by $U = (c - c_{sat}^S)/c_{sat}^S$, where $c_{sat}^S(T)$ is the saturation number density of vapor of ice at temperature $T$. The space/time evolution equations are written as~\cite{demange2017phase,younsi2016anisotropy,jeong2001phase,beckermann1999modeling}:
\begin{multline}
    \tau_0 a_s^2(\textbf{n}) \frac{\partial \phi}{\partial t} =\\ W_0^2  \bm{\nabla} \cdot \left(a_s^2(\textbf{n})\right) \bm{\nabla} \phi +  W_0^2 \bm{\nabla} \cdot \left (|\bm{\nabla} \phi|^2  \frac{\partial[a(\textbf{n})^2]}{\partial \bm{\nabla} \phi}\right )\\
    + (\phi - \phi^3) + \lambda U (1 - \phi^2)^2 ,
  \label{a}
\end{multline}
and:
\begin{equation}
    \frac{\partial U}{\partial t} + \left( \frac{1-\phi}{2}\right) \bm{u} \cdot \bm{\nabla} U  =D \bm{\nabla} \cdot \left(q(\phi) \bm{\nabla} U \right) - \frac{L_{sat}}{2} \frac{\partial \phi}{\partial t},
     \label{b}
\end{equation}
defining for the later analysis $\tau = \tau_0 a_s^2(\textbf{n})$. Here, $\tau_0$ denotes the characteristic time and $W_0$ denotes the characteristic width of the diffuse interfaces. In Eq.~\ref{a}, the coefficient $\lambda = \frac{15L^2}{16Hc_pT_m}$ describes the strength of the coupling between the phase-field and the species field. $1/\lambda$ is a dimensionless measure of the barrier height $H$ of the double-well potential. The latent heat of melting is shown as $L$. The specific heat capacity $c_p$ is assumed to be the same in the two phases (symmetric model). $\textbf{n} = - \frac{\bm{\nabla} \phi}{\left| \bm{\nabla} \phi \right|}$ is the unit vector normal to the crystal interface --pointing from solid to the fluid. $a_s(\textbf{n})$ is the surface tension anisotropy function which here, in the context of the snowflake growth studies, is defined as:
\begin{equation}
    a_s(\textbf{n}) = 1 + \epsilon_{xy} \cos(6 \theta),
\end{equation}
\newabbreviation{rhs}{RHS}{right hand side}
\newabbreviation{lhs}{LHS}{left hand side}
where $\theta = \arctan(n_y/n_x)$. $\epsilon_{xy}$ is a numerical parameter characterizing the anisotropy strength. The second and third terms on the \gls{rhs} of Eq.~\ref{a} control the anisotropic growth of dendrites in snowflakes. $\phi - \phi^3$ is the derivative of the double-well potential. The last term in Eq.~\ref{a} is a source term accounting for the coupling between the supersaturation $U$ and the order parameter $\phi$. $(1 - \phi^2)^2$ is an interpolation function minimizing the bulk potential at $\phi = \pm 1$.\\
In Eq.~\ref{b}, $\bm{u}$ denotes the fluid local velocity while $q(\phi) = (1 - \phi)$ is a function canceling out diffusion within the solid.  As such solute transport is assumed to take place only within the fluid phase (one-side model). The parameter $D$ is the diffusion coefficient of water vapor in air as applied in Section 3.2.1. $L_{sat}$ describes the absorption rate of water vapor at the interface of the snow crystals.\\
The thermal capillary length is defined as:
\begin{equation}
    d_0 = \frac{\gamma_0 T_M c_p}{L^2},
    \label{d0}
\end{equation}
where the excess isotropic free energy of the solid-liquid interface $\gamma_0$ is given by:
\begin{equation}
    \gamma_0 = I W_0 H,
\end{equation}
where $W_0$ is the interface thickness, $H$ is the barrier height of the double-well potential and $I = 2 \sqrt{2}/3$. Hence, 
 the matched asymptotic expansions provide a relation between the phase-field and sharp-interface parameters given by:
\begin{equation}
    d_0 = a_1 \frac{W_0}{\lambda}.
\end{equation}
The expression for the kinetic coefficient is:
\begin{equation}
    \beta = a_1 \left( \frac{\tau_0}{W_0 \lambda} - a_2 \frac{W_0}{D}\right)
    \label{s}
\end{equation}
where $a_2$ is a constant and depends on the choice of the function $g(\phi)$. In this paper, the standard form  $g(\phi) = \phi - 2\phi^3/3 + \phi^5/5$  will be used. For the model, $a_1 \approx 0.8839$ and $a_2\approx 0.6267$. These relations make it possible to choose phase-field parameters for prescribed values of the capillary length (surface energy) and the interface mobility (interface kinetic coefficient). Note that the interface width $W_0$ is a parameter that can be freely chosen in this formulation; the asymptotic analysis remains valid as long as $W_0$ remains much smaller than any length scale present in the sharp-interface solution of the considered problem \cite{karma1998quantitative,echebarria2004quantitative,folch2005publisher}. \\
It follows from Eq.(\ref{s}) that $\beta$ can be made to vanish provided that $\tau$ is chosen to be:
\begin{equation}
    \tau = \tau_0 \left[A(n) ^2 \right].
    \label{t}
\end{equation}
Therefore,
\begin{equation}
    \tau_0 = a_2 \lambda \frac{W_0^2}{D}.
\end{equation}
The results should be independent of $\lambda$ when they are converged. Decreasing $\lambda$ corresponds physically to decreasing the interface width while increasing at the same time the height of the double-well potential so as to keep the surface energy and hence $d_0$ fixed.\\
The physical dimensions of $W_0$, $\tau_0$ and $\lambda$ are respectively $[W_0] \equiv [\mathcal{L}]$,  $[\tau_0] \equiv [\mathcal{T}]$, $[\lambda] \equiv [-]$,  $[D] \equiv [\mathcal{L}^2/\mathcal{T}]$ where $[\mathcal{L}]$ indicates the length dimension and $[\mathcal{T}]$
indicates the time dimension.
\subsection{Lattice Boltzmann formulation}
\subsubsection{Flow field solver}
The target system of equations describing the flow field behavior (i.e. incompressible \gls{ns}-continuity equations) are modeled using the classical isothermal \gls{lb} formulation consisting of the now-famous stream-collide operators:
\begin{equation}
    f_\alpha \left( \bm{x}+\bm{c}_\alpha \delta_t, t+\delta_t\right) - f_\alpha \left( \bm{x}, t\right) = \delta_t \Omega_\alpha\left( \bm{x}, t\right) + \delta_t\bm{F},
\end{equation}
where $\bm{F}$ is the external force. Here, the external force $\bm{F}$ modeling interaction with the solid phase is given as~\cite{beckermann1999modeling}:
\begin{equation}
    F = -\frac{h\eta_f (1+\phi)^2 (1-\phi)\bm{u}}{4W_0^2}
\end{equation}
where $h$ is a dimensionless constant, here $h = 2.757$. Due to the absence of fluid velocity in snow crystals, the fluid velocity $\bm{u}$ is updated as:
\begin{equation}
    \bm{u^*} = \frac{(1-\phi)}{2}\bm{u},
\end{equation}
where the updated fluid velocity $\bm{u^*}$ is taken into the momentum equation. The above friction term acts as a distributed momentum sink that gradually forces the liquid velocity to zero as $\phi \rightarrow 1$.
The collision operator $\Omega_\alpha$ follows the linear Batnagar-Gross-Krook approximation:
\begin{equation}
    \Omega_\alpha = \frac{1}{\tau}\left[f^{(eq)}_\alpha  - f_\alpha\right],
\end{equation}
\newabbreviation{edf}{EDF}{equilibrium distribution function}
where $f_\alpha^{(eq)}$ is the discrete isothermal \gls{edf} defined as:
\begin{equation}
    f_\alpha^{(eq)} = \rho\sum_i \frac{1}{2 c_s^2} a^{(eq)}_i(\bm{u}):\mathcal{H}_{i}(\bm{c}_\alpha),
\end{equation}
where $a^{(eq)}_i$ and $\mathcal{H}_{i}(\bm{c}_\alpha)$ are the corresponding multivariate Hermite coefficients and polynomials of order $i$, with $c_s$ the lattice sound speed corresponding to the speed of sound at the stencil reference temperature~\cite{hosseini2020compressibility}. Further information on the expansion along with detailed expressions of the \gls{edf} can be found in~\cite{shan2006kinetic,hosseini2019extensive,hosseini2020development}. The relaxation time $\tau$ is tied to the fluid kinematic viscosity as:
\begin{equation}
    \tau = \frac{\nu}{c_s^2} + \frac{\delta_t}{2}.
\end{equation}
It must be noted that conserved variables, {i.e.} density and momentum are defined as moments of the discrete distribution function:
\begin{equation}
    \rho = \sum_\alpha f_\alpha,
\end{equation}
\begin{equation}
    \rho \bm{u} = \sum_\alpha \bm{c}_\alpha f_\alpha.
\end{equation}
\subsubsection{Advection-diffusion-reaction solver for supersaturation field}
The space/time evolution equation of the water vapour supersaturation field is modeled using an advection-diffusion-reaction \gls{lb}-based discrete kinetic formulation defined as\cite{ponce1993lattice}:
\begin{equation}
    g_\alpha \left( \bm{x}+\bm{c}_\alpha \delta_t, t+\delta_t\right) - g_\alpha \left( \bm{x}, t\right) = \delta_t \Omega_\alpha\left( \bm{x}, t\right) + \delta_t \dot{\omega}_\alpha,
\end{equation}
where $\dot{\omega}_\alpha$ is the source term defined as:
\begin{equation}
    \dot{\omega}_\alpha = - w_\alpha \frac{L_{sat}}{2}\frac{\partial \phi}{\partial t}.
\end{equation}
The collision operator $\Omega_\alpha$ for the supersaturation field is:
\begin{equation}
    \Omega_\alpha = \frac{1}{\tau_g}\left[g^{(eq)}_\alpha  - g_\alpha\right].
\end{equation}
where $g_\alpha^{(eq)}$ is the \gls{edf} defined as:
\begin{equation}
    g^{(eq)}_\alpha = \omega_\alpha U\left[ 1 + \frac{\bm{c}_\alpha \cdot \bm{u} }{c_s^2} \right].
\end{equation}
The supersaturation is computed locally as the zeroth-order moment of $g_\alpha$:
\begin{equation}
    U = \sum_\alpha g_\alpha,
\end{equation}
and the relaxation coefficient is tied to the water vapour diffusion coefficient~\cite{huber2010lattice} in air:
\begin{equation}
    \tau = \frac{D q(\phi)}{c_s^2} + \frac{\delta_t}{2}.
\end{equation}

\subsubsection{Solver for phase-field equation}
The phase-field equation is modeled using a modified \gls{lb} scheme defined as~\cite{walsh2010macroscale,cartalade2016lattice}:
\begin{multline}
  a_s^2(\bm{n}) h_\alpha(\bm{x} + \bm{c}_\alpha \delta x, t + \delta t) =\\ h_\alpha(\bm{x},t) -  \left( 1 - a_s^2(\bm{n})  \right ) h_\alpha(\bm{x} + \bm{c}_\alpha \delta x, t) - \\
   \frac{1}{\eta_\phi (\bm{x},t) }
  \left [ h_\alpha(\bm{x},t) - h_\alpha^{eq}(\bm{x},t) \right]  + \omega_\alpha Q_\phi (\bm{x},t)\frac{\delta t}{\tau_0},
  \label{d}
\end{multline}
where the scalar function $Q_\phi$ is the source term of the phase-field defined by:
\begin{equation}
    Q_\alpha = (\phi - \phi^3) + \lambda U (1 - \phi^2)^2,
\end{equation}
while the \gls{edf} $h_i^{eq}$ is defined as:
\begin{equation}
    h_\alpha^{eq} = \omega_\alpha \left( \phi - \frac{1}{c_s^2} \bm{c}_\alpha \cdot \frac{W_0^2}{\tau_0} |\bm{\nabla} \phi|^2 \frac{\partial (a_s(\bm{n})^2}{\partial \bm{\nabla} \phi} \frac{\delta t}{\delta x} \right).
    \label{e}
\end{equation}
The local value of the order parameter $\phi$ is  calculated as:
\begin{equation}
    \phi = \sum_{\alpha} h_\alpha,
\end{equation}
The relaxation time $\eta_\phi$ is a function of position and time and must be updated at each time-step as:
\begin{equation}
  \eta_\phi = \frac{1}{c_s^2}a_s^2(\bm{n})\frac{W_0^2}{\tau_0} + \frac{\delta_t}{2},  
\end{equation}
\section{Simulations and numerical studies}
For the sake of clarity, this part is organized into two different subsections. In the first subsection results from simulations of generic systems, intended as validation are presented and discussed. The results for snowflakes in the plate regime are presented in the second subsection.
\subsection{Validation studies}
\subsubsection{Effect of directional derivatives of gradients in LB scheme}
In order to calculate the derivatives of $a_s(\bm{n})$ with respect to $\partial_x \phi$ and $\partial_y \phi$ involved in the second term in Eq.~\ref{a}, different choices are available. Overall, one can either evaluate the gradient using classical finite-difference approximations, {e.g.} the central difference second-order formulation: $\partial_x \phi \simeq (\phi_{i+1,j} - \phi_{i-1,j})/2 \delta x $ and $\partial_y \phi \simeq (\phi_{i,j+1} - \phi_{i,j-1})/2 \delta x$, where $i$ and $j$ are the indexes of the coordinates $x$ and $y$, or the method based on the directional derivatives with higher-order isotropy \cite{chen2014critical}. The isotropic finite-difference approximations to the first-order derivative can be computed as:
\begin{equation}
\nabla \phi = \frac{1}{c_s^2}\sum_{\alpha=0}^Q w_\alpha\left( |c_\alpha|^2 \right) \phi \left( x + c_\alpha \right) c_\alpha 
\end{equation}
where $w_\alpha\left( |c_\alpha|^2 \right)$ are the weights associated to each layer of neighboring nodes maximizing isotropy~\cite{shan2006analysis}. These weights are summarized in table~\ref{t1}. Given the importance of this directional derivative in the growth dynamics of the solid phase, the choice of the approximation will be briefly discussed here.
 \begin{table}[!htp]
 \centering
 \setlength{\tabcolsep}{1.0mm}{
 \caption{Weights for 4th, 6th and 8th order isotropic tensors in two dimensions~\cite{sbragaglia2007generalized,shan2006analysis}}
\begin{tabular}{ccccccccc}
\hline
Tensor& $w(1)$  &  $w(2)$ &$w(3)$ & $w(4)$&$w(5)$ &$w(6)$ &$w(7)$ &$w(8)$ \\
 \hline
E4 &   1/3   &   1/12   &   -  &  - & - & - & - & -  \\
E6 &    4/15     &   1/10  &  -   &  1/120    &  - & -& -&-   \\
E8 &  4/21   &  4/45   &  - & 1/60 & 2/315 & -&-&1/5040 \\
\hline  
\label{t1}
\end{tabular}}
\end{table}
\begin{figure}[!htbp]
	\centering
	\includegraphics[width=0.4\textwidth]{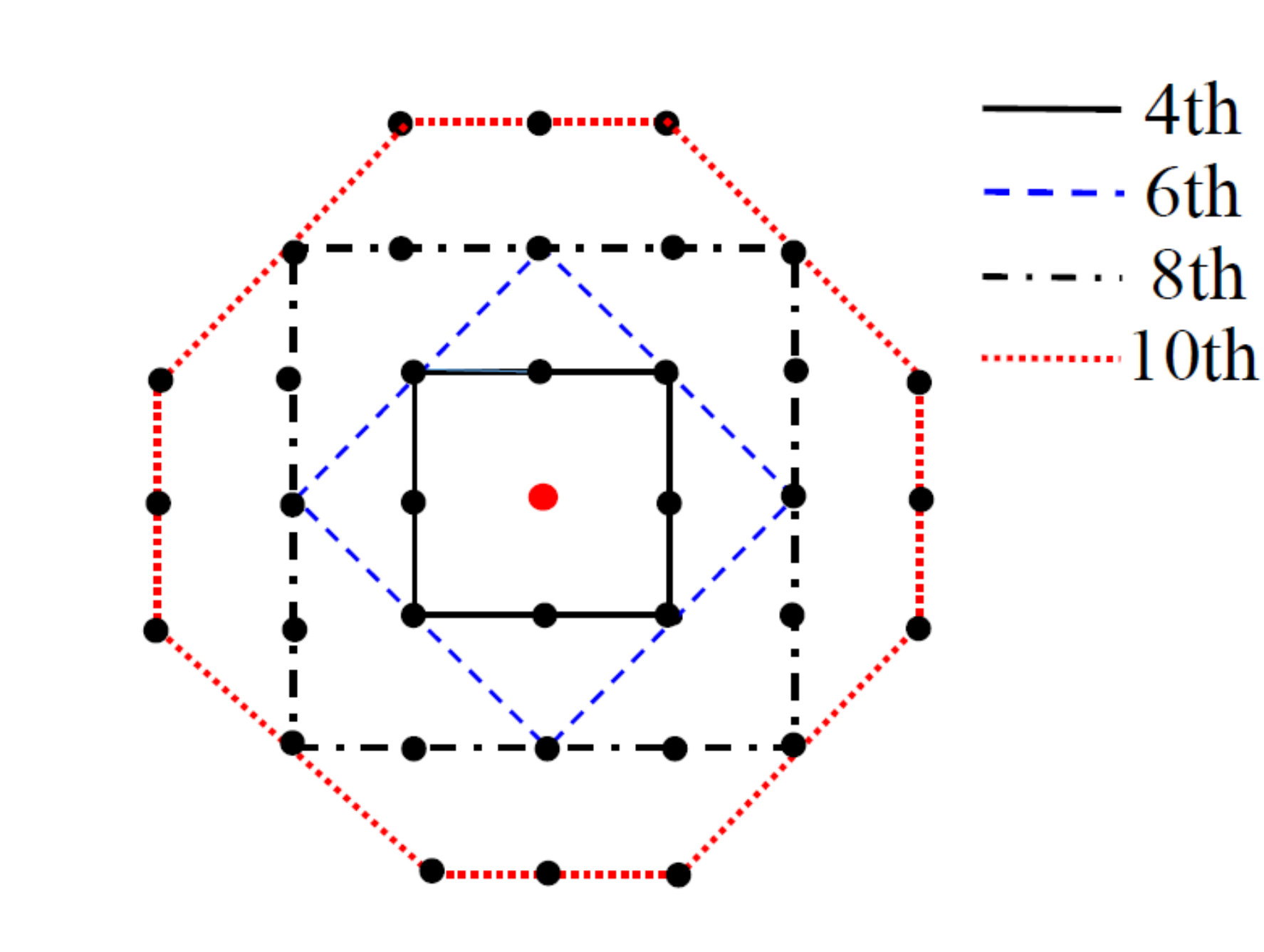}
	\caption{The grid points identifying the set of
		velocity fields for a 2d case from~\cite{sbragaglia2007generalized,shan2006analysis}. With reference to the weights reported
		in Table~\ref{t1}, different degrees of isotropy can be achieved: fourth
		order [up to $\omega$(2)], sixth order [up to $\omega$(4)], eighth order [up to $\omega$(8)].}
	\label{iso_order}
\end{figure}
To that end we consider a generic growing crystal with hexagonal symmetry driven by the temperature field. For all simulations a domain of size $800\times 800$, with $\delta_x = \delta_y = 0.01 $ and $ \delta_t = 1.5 \times 10^{-5}$ is considered. Furthermore, $\tau_0 = 1.5625 \times 10^{-4}$, $W_0 = 0.0125$, $ \kappa = 1$, the undercooling is $ \Delta = 0.3$, $\lambda = 10$, $\varepsilon_s = 0.05$. The circular seed is initialized at the center of the domain $\bm{X_c}= (400,400)$ with a radius of $R_s= 10$ lattice units.\\
\begin{figure*}[!ht]
	\centering
	\hspace{-0.4\textwidth}
	\begin{subfigure}{0.3\textwidth}
		\includegraphics{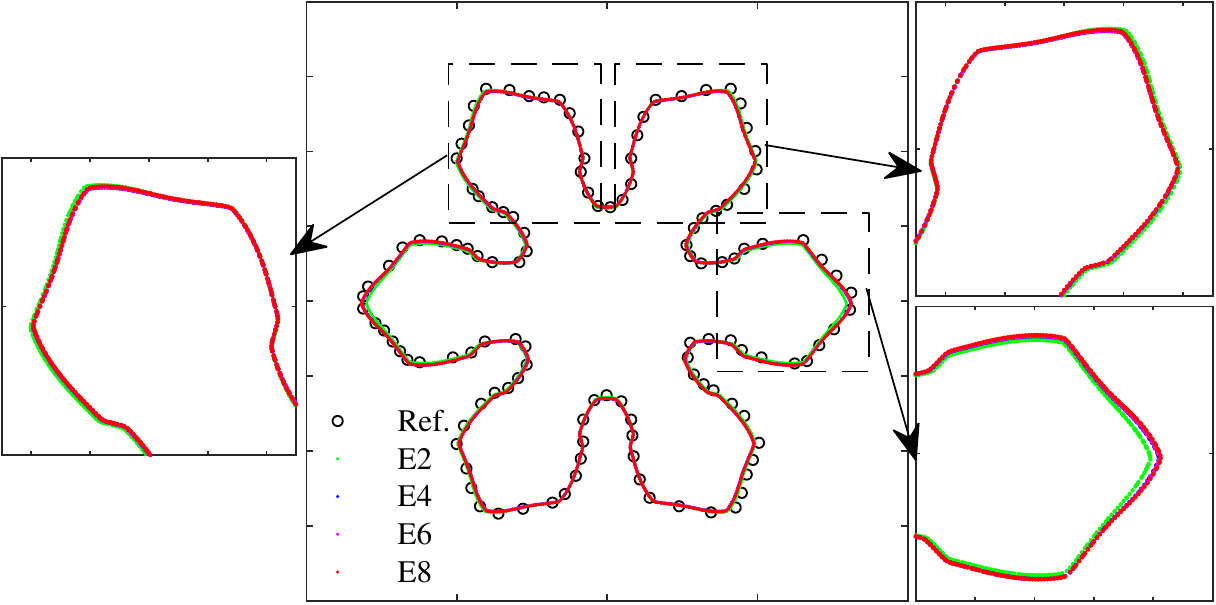}
	\end{subfigure}
	\caption{Iso-contours of $\phi =0$ at time $t = 1 \times 10^{5} \delta_t$. Black circles are reference results from~\cite{younsi2016anisotropy}.}
	\label{dd}
\end{figure*}
The solid boundaries, $\phi = 0$,  as obtained using different approximations at $t = 10^5 \delta_t $ are shown in Fig.~\ref{dd}. It could be seen that results obtained with a second-order finite difference method (E2) do not match well with the reference. As a function of their orientation relative to the main axes, the dendrites are slightly different from each other. When the gradients are calculated with higher order isotropic formulations these non-physical anisotropic effects are considerably reduced. As such a minimum isotropy of order 4, {i.e.} the E4 stencil, is necessary for a six-tip crystal simulation.
\subsubsection{Validation for crystal growth}
To validate the model and subsequent implementation, we use the generic system with four-fold symmetry studied in~\cite{younsi2016anisotropy,cartalade2016lattice}. To provide both qualitative and quantitative proof, we compare the shape of the dendrites and the evolution of the tip velocity.\\
Initially, a circular seed of radius $R_s = 10 \delta_x$ is placed at the centre of the square domain. The interface thickness $W_0$ and the characteristic time $\tau_0$ are $W_0 = \tau_0 = 1$. The grid-size is set to $\delta_x / W_0 =0.4$~\cite{karma1998quantitative} while the time-step size is $\delta_t = 0.008$. The capillary length $d_0$ and the kinetic coefficient $\beta$ are given by~\cite{karma1998quantitative}: $d_0 = a_1 W_0/\lambda$ and $\beta = a_1 (\tau_0 /\lambda W_0 - a_2 W_0 /\kappa)$ where $a_1 = 0.8839$ and $a_2 = 0.6267$. In this benchmark, we choose the parameters $\kappa = 4$, $\lambda = \frac{\tau_0 \kappa}{a_2 W_0^2} = 6.3826$ with $\beta = 0$. The anisotropic strength is $\epsilon_s = 0.05$. The initial supersaturations are $U_0 = 0.3$ and $0.55$. For $U_0 = 0.3$, we use a grid of size $1000^2$ nodes while for the latter the domain size is reduced to $500^2$ nodes. The growth velocity of the crystal tips $\tilde{\bm{V_p}}$ is made dimensionless as  $\tilde{\bm{V_p}} = \bm{V_p} d_0/D$. The non-dimensional position $\tilde{\bm{x}}$ is defined as ($\tilde{\bm{x}} = \bm{x}/\delta x$) and the reduced time $T$ as ($T = t/\tau_0$).\\
\begin{figure*}[!ht]
	\centering
	\hspace{-0.4\textwidth}
	\begin{subfigure}{0.3\textwidth}
		\includegraphics{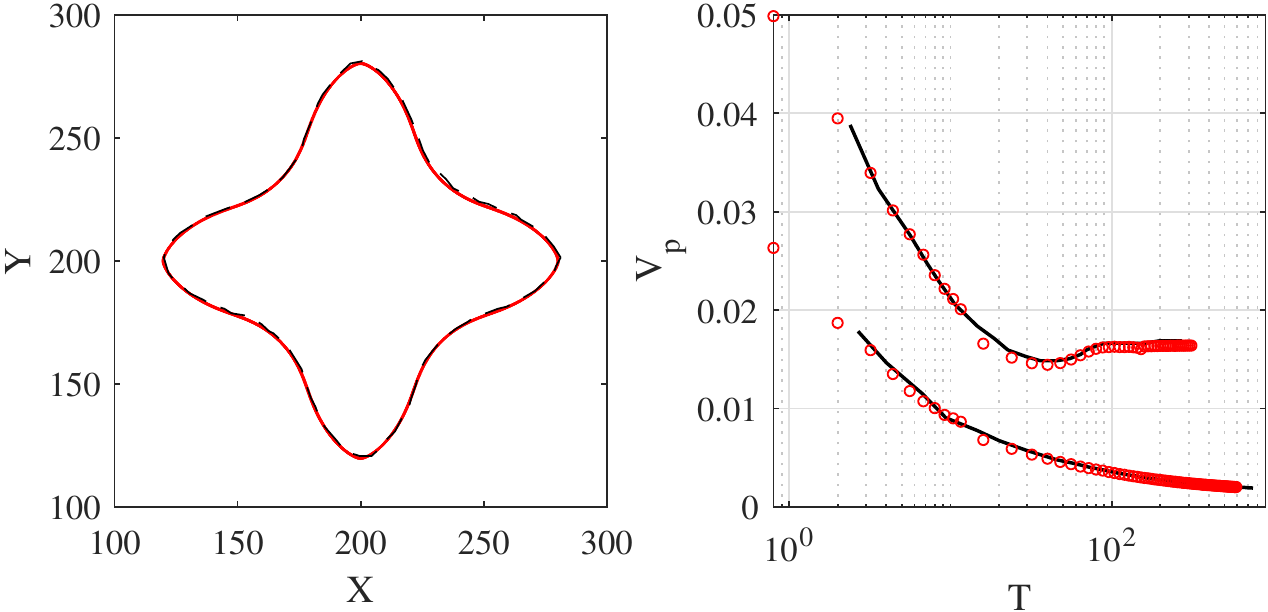}
	\end{subfigure}
	\caption{(left) $\phi=0$ iso-contours for $U_0 = 0.3$ at $t= 1.3 \times 10^5 \delta_t$. Red dots are from the present study while dashed black lines are from~\cite{cartalade2016lattice}. (right) Dimensionless tip velocity $V_p$ as a function of time (in units of $\tau_0$) for (lower curve/symbols) $U_0 = 0.3$ and (upper curve/symbols) $U_0 = 0.55$. The red circles are from the present study, while plain black lines are extracted from~\cite{cartalade2016lattice}.}
	\label{pt2}
\end{figure*}
The obtained results are shown in Fig.~\ref{pt2}. As shown there, the data obtained from the present work closely follows those reported in~\cite{cartalade2016lattice}.
\subsubsection{Validation of flow/solid coupling}
\begin{figure*}[!ht]
	\centering
	\hspace{-0.35\textwidth}
	\begin{subfigure}{0.3\textwidth}
		\includegraphics{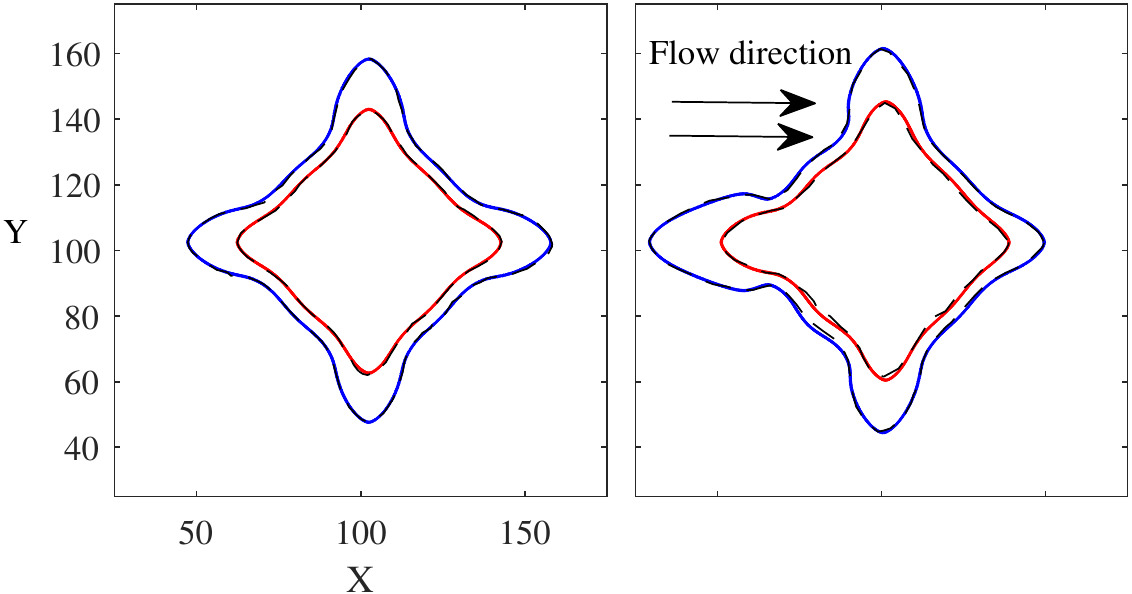}
	\end{subfigure}
	\caption{Computed phase-field contours from the dendritic growth in 2-D (left) without and (right) with flow at two different times. Red and blue symbols are \gls{lb} results at $t=72$ and $104$ while black dashed lines are corresponding reference data.}
	\label{fp}
\end{figure*}

\begin{figure}[!ht]
	\centering
	\hspace{-0.25\textwidth}
	\begin{subfigure}{0.25\textwidth}
		\includegraphics{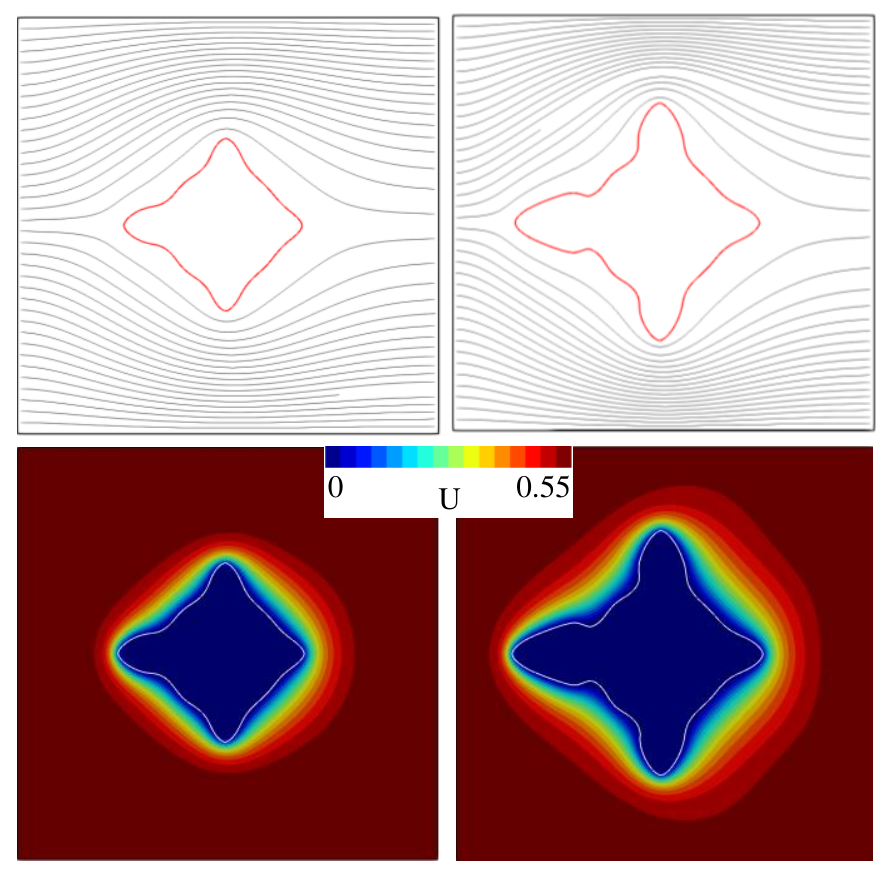}
	\end{subfigure}
	\caption{(top row) Velocity field streamlines and (bottom row) concentration fields at two different times: (left) $T=72$ and (right) $104$.}
	\label{ftp1}
\end{figure}

In order to validate the coupling between the flow field and other solvers, we model the 2-D case presented in~\cite{beckermann1999modeling,jeong2001phase}, and solved there via an adaptive finite-elements solver. The computational domain is a box of length $L=204.8$, the grid-size $\delta_x = 0.4$ and the time-step unit $\delta_t = 0.008$. We also set the interface thickness to $W_0 = 1$, the characteristic time to $\tau_0 = 1$, the coupling strength to $\lambda = 6.383$ and the anisotropy strength to $\epsilon_s = 0.05$. The kinematic viscosity of the flow field is $\mu_f = 92.4$. The flow enters from the left side of the box with a fixed inlet velocity $u_x = 1.0$. The \gls{rhs} boundary is set to zero-gradient. The top and bottom sides of the box are set to periodic.\\
The evolution of all dendrite tips without/with velocity are plotted in Fig.~\ref{fp}. 
It can be observed that the phase-field contours are in excellent agreement with the reference. To better illustrate the interaction of the flow field with the growing solid, the streamlines and species concentration field at two different times are shown in Fig.~\ref{ftp1}. As expected, one can readily see the non-isotropic distribution of the concentration field around the growing seed caused by the flow field. The incoming velocity induces a higher concentration gradient around the tip on the \gls{lhs} causing it to grow faster than its counter-part in the opposite direction.
\subsection{Faceted and dendritic snowflake growth}
In this subsection, we first study the plate growth regime of snowflakes as a function of the supersaturation to showcase the ability of the proposed model to capture the wide variety of habits exhibited by snowflakes. Then, we briefly look at the effect of forced convection on the growth of snowflakes.
\subsubsection{Ice crystal habit in thin-plate regime as a function of temperature and supersaturation}
\begin{figure*}                        
	\begin{subfigure}{0.3\textwidth}
	\includegraphics[width=3.4\textwidth]{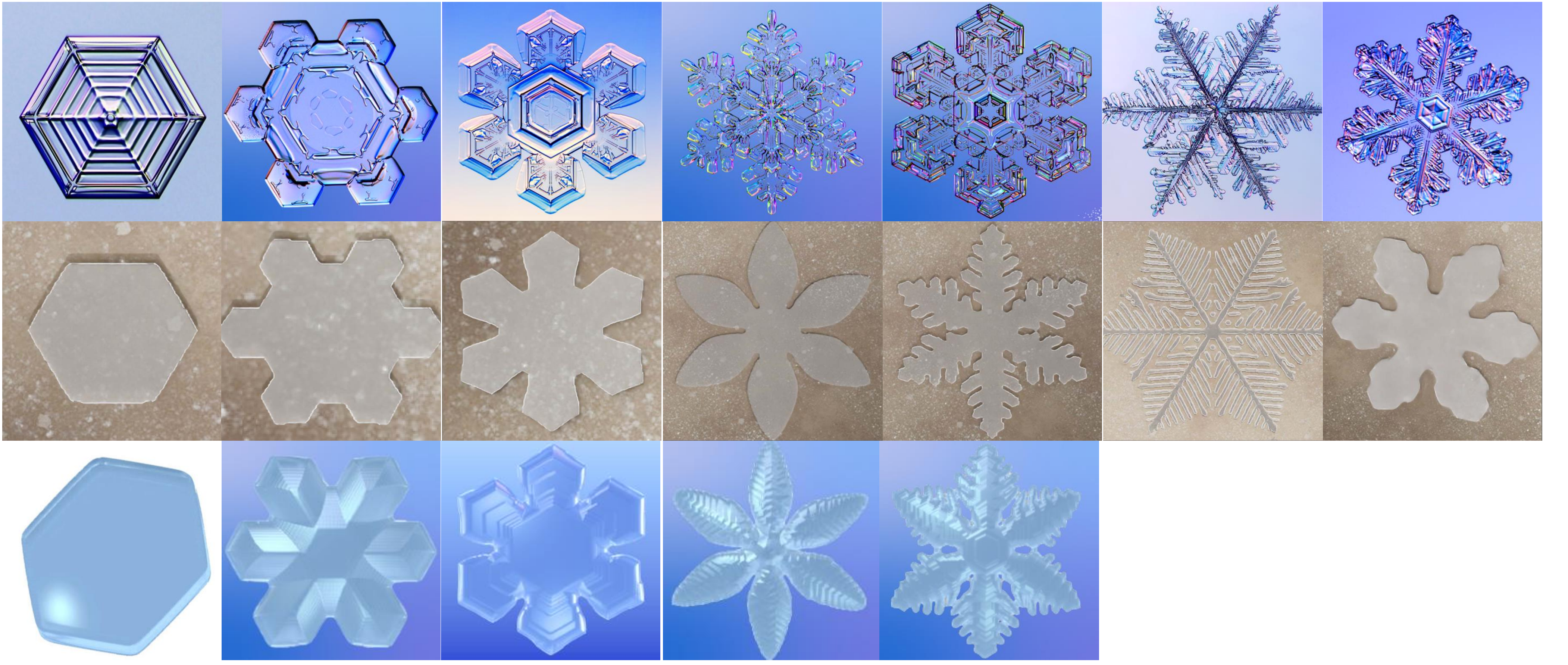}
	\end{subfigure}
	\caption{The morphology numbers are from 1 to 7 (from left to right) in this figure, as indicated in table~\ref{t2}. Comparison between (top) real snowflakes photographs taken from Libbrecht's experiments~\cite{RN16}, (middle) our phase-field simulations, and (bottom) the simulations from~\cite{demange2017phase}.}
	\label{ff}
\end{figure*}

As stated earlier, the focus of the present study is on the plate growth regime. According to the Nakaya diagram, the widest variety of habits and instabilities in this regime can be observed at $-16^{\circ}$C where the saturated vapor density of ice is around $\rho_{sat}^I = 1.125\hbox{g}/\hbox{m}^3$ and the saturated vapor density of water is around $\rho_{sat}^W = 1.53\hbox{g}/\hbox{m}^3$~\cite{cartalade2016lattice}. The excess density over vapor-water equilibrium is $\Delta \rho = \rho - \rho_{sat}^W$ which is exhibited in Fig.~\ref{1} by y-axis, where $\rho_{sat}^W$ is the satuaration vapor density of water. The supersaturation is given by $U = (c-c_{sat}^I)/c_{sat}^I$, where $c_{sat}^I$ is the saturation number density of vapor of ice at temperature $T$. Using $\rho = m_{H_2o}c$, where $m_{H_2o}$ is the mass of a molecule of water, it could be written $ U = (\Delta \rho + \rho_{sat}^W - \rho_{sat}^I)/\rho_{sat}^I$ and different snowflake morphologies are obtained by changing the initial excess vapor density $\Delta \rho$. According to~\cite{yoshizaki2012precise}, the melting temperature of the snow crystals is $T_m= 276.9\hbox{K}$, heat capacity $c_p = 4.23\times 10^6 \hbox{J} \hbox{m}^{-3} \hbox{K}^{-1}$, diffusion coefficient of the vapor $D=1.17\times 10^{-7} \hbox{m}^2\hbox{s}^{-1}$ which is illusrated in Eq.~\ref{b}, surface tension $\gamma = 2.845\times10^{-2}\hbox{J} \hbox{m}^{-2}$, and latent heat $L=1.12\times10^8\hbox{J} \hbox{m}^{-3}$. The capillary length is computed to be $d_0 \simeq 2$~nm. The coupling parameter is set to $\lambda = 3$~\cite{ramirez2004phase}.\\
The simulation parameters are chosen as: $W_0=1.25\delta_x$ and $\tau_0 = 20 \delta_t$. The simulations were conducted in a 400$\times$400 box. The anistropy strength was set to $\varepsilon_{xy}=0.05$. To produce different snowflake morphologies, the following parameters were varied: initial supersaturation $u_0$, depletion rate $L_{sat}$, initial density $\rho_0$ and the excess density $\Delta \rho^W$. The corresponding values are listed in table~\ref{t2}. 
 \renewcommand\arraystretch{1.4}
 \begin{table}[!htbp]
 \centering
 \setlength{\tabcolsep}{1.4mm}{
 \caption{Parameters for different morphologies of snowflakes}
\begin{tabular}{cccccc}
\hline
 Nr.&shapes& $U_0[-]$  &  $L_{sat} [-]$ & $\rho_0 [\hbox{g}/\hbox{m}^3]$ & $\Delta \rho^W [\hbox{g}/m^3]$\\
 \hline
1&solid plate        &   0.4   &   2.0   &   1.575  &  0.045     \\
2&stellar plate I    &   0.5   &   1.8   &   1.6875 &  0.1575    \\
3&sectored plate     &   0.5   &   1.6   &   1.6875 &  0.1575    \\
4&stars              &   0.6   &   1.0   &   1.8    &  0.27      \\
5&fern dendrite      &   0.8   &   1.6   &   2.025  &  0.495     \\
6&fernlike stellar   &   0.7   &   1.6   &   1.9125 & 0.383      \\
7&stellar plate II   &   0.5   &   1.2   &   1.6875 &  0.1575    \\
\hline  
\label{t2}
\end{tabular}}
\end{table}
The results obtained for the conditions listed in table~\ref{t2} are shown in Fig.~\ref{ff} along with corresponding numerical and experimental data from~\cite{demange2017phase} and \cite{RN16}.
The primary habit of the crystal (six-fold symmetry) is dictated by the anisotropy function (and the microscopic crystallographic structure). At lower supersaturation values where the adsorption rate is slow, the surface diffusion process characteristic time is smaller and therefore dominates over surface adsorption. More explicitly it means that the adsorbed water molecules have enough time to propagate on the crystal surface and find the points with the lowest potential (dictated by the molecules' arrangement in the crystal lattice). Furthermore, given the low growth rate and gradients, the surface is not subject to branching instabilities. As the supersaturation goes up, the larger adsorption rate at the sharper parts of the interface (regions with the highest curvatures and consequently highest surface area) result in the formation of six thick branches (usually referred to as primary branches). In the lower supersaturation regimes these primary branches have a faceted structure following the symmetry of the crystal. As the concentration goes further up, the branches get thinner and rougher (the straight faces tend to disappear); this eventually produces secondary instabilities and branches going towards a somewhat fractal structure. All the obtained crystal habits, are in excellent agreement with not only numerical simulations from~\cite{demange2017phase} but also experimental data from~\cite{RN16}. Further comparing the different crystal habits to Nakaya's diagram, it can be concluded that the proposed model correctly predicts the behavior of the crystal in the platelet regime. The next part will focus on the effects of ventilation on the evolution of the crystal habit.
\subsubsection{Ventilation effect on snowflakes}
\begin{figure}[htb]                       
	\begin{subfigure}{0.3\textwidth}	
	\includegraphics[width=1.5\textwidth]{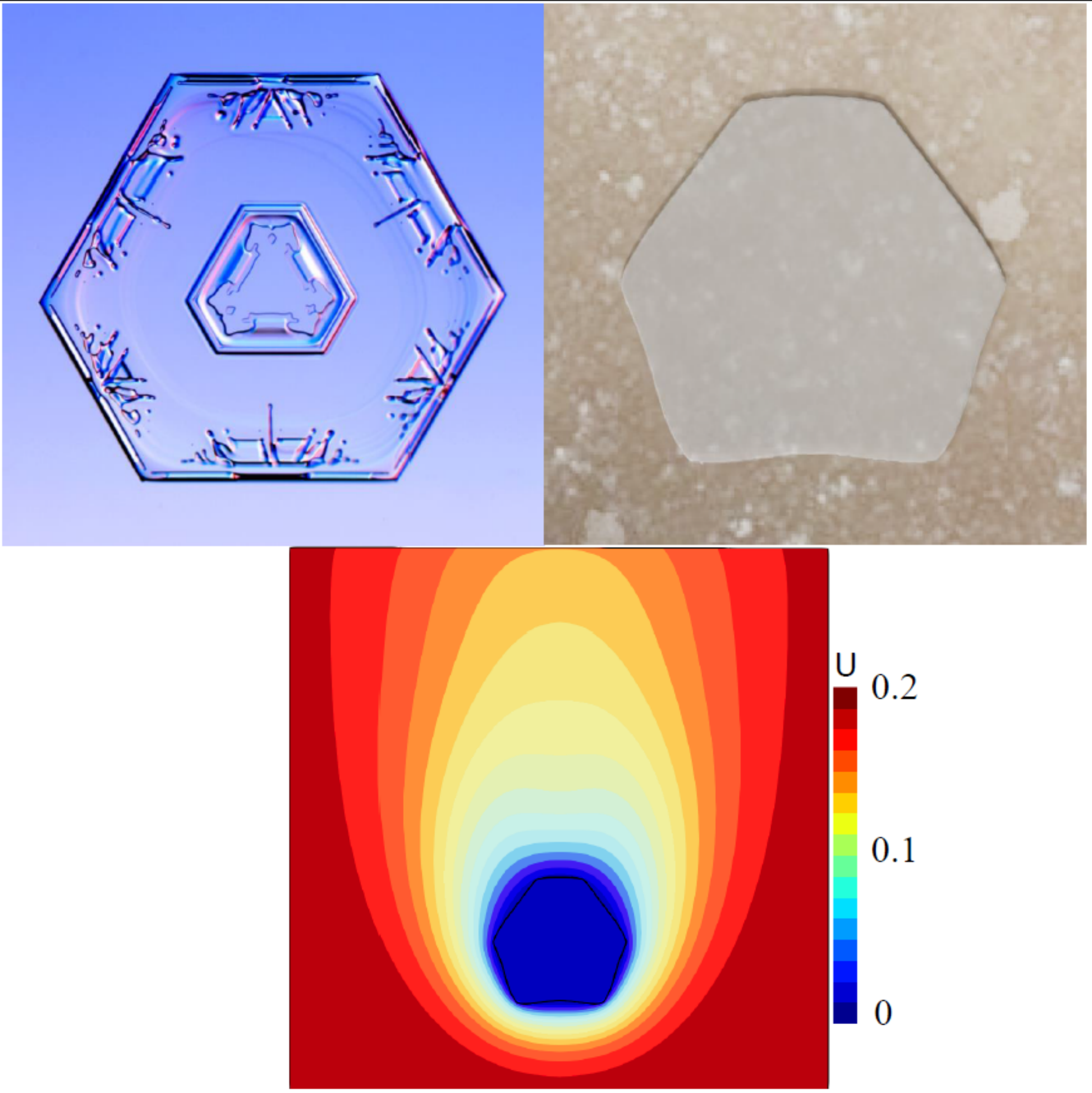}
	\end{subfigure}
	\caption{The morphology of the asymmetrical hexagonal snowflakes at $t = 7500$ in units of $\tau_0$. Top left: experimental image, top right: our \gls{lb} simulation, and bottom: associated supersaturation field.}
	\label{ve}
\end{figure}

Growth of snowflakes under forced convection is a topic of interest, as for example falling flakes are usually subject to ventilation. The dynamics of snowflake growth under ventilation effects are not very well-documented. Libbrecht and Arnold~\cite{libbrecht2008physically} worked on an aerodynamic model to show the growth and appearance of new habits such as the triangular snow crystals both in nature and in laboratory settings~\cite{libbrecht2009aerodynamic}. Furthermore, the anisotropy induced by the flow field can, when pronounced, cause different regions on the surface of the crystal to grow in different regimes. The present section will focus on recovery of these effects and qualitative validation with experimental observations.\\
For the first configuration, the domain has a size of $1600\times 1600$ grids, the initial seed radius is $R=25\delta x$ and the initial vapor density $\rho_0 = 1.364$g/m$^3$ (supersaturation $U_0=0.2$). The grid size is set to $\delta_x= 4.8 \times 10^{-6}$~m while the time-step is $\delta_t = 2\times 10^{-7}$~s. The coupling strength is $\lambda = 3$. The flow blows from the bottom to the top along the $y$-axis. The inlet velocity is set to $\bm{u}_y= 0.12 \hbox{m}/\hbox{s}$ and the kinetic viscosity to $\nu_f = 1.152\times 10^{-6}\hbox{m}^2/\hbox{s}$. The outlet is modeled using a Neumann condition, while along the $x$-axis periodic boundary conditions are used. The resulting snowflake morphology is shown in Fig.~\ref{ve}. 
It can be clearly observed that the crystal growth on the side facing the incoming flow is higher than its neighboring sides. Furthermore, its opposite side is growing slower than its neighbour. Both of these non-symmetrical growth rates push the habit towards, first a non-symmetrical hexagon and then a triangular shape, in agreement with experimental observations. 
\begin{figure}[!ht]
	\centering
	\hspace{-0.25\textwidth}
	\begin{subfigure}{0.25\textwidth}
		\includegraphics{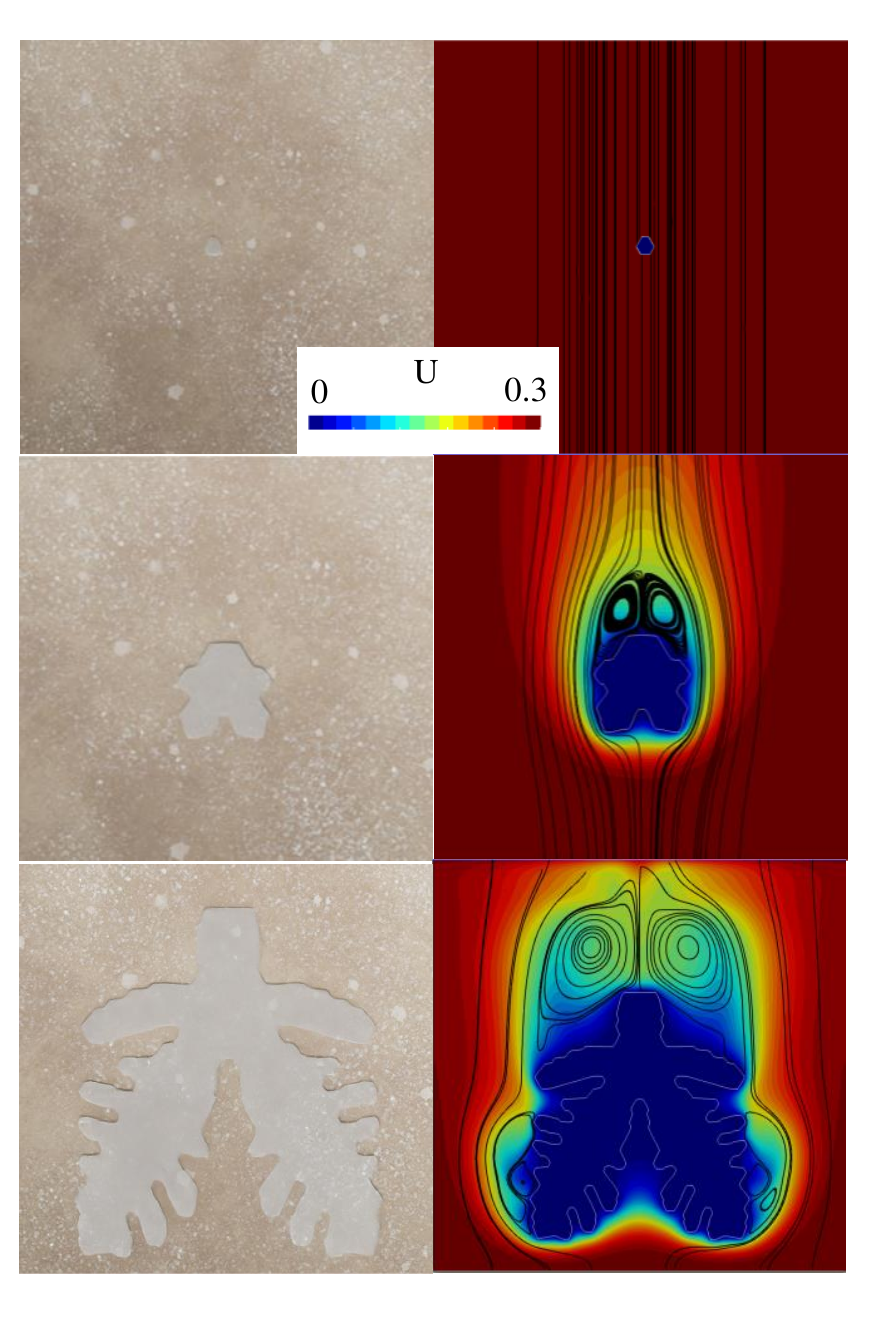}
	\end{subfigure}
	\caption{(left) morphology of the snowflakes with ventilation effects and (right) velocity field streamlines and supersaturation fields at (from top to bottom) $t$ = 0, 4000 and 8000 in units of $\tau_0$.}
	\label{v}
\end{figure}
To further put the effect of hydrodynamics into perspective we also consider another test-case with the same configuration as the previous one, however with a larger initial supersaturation. The initial vapor density is set to $\rho_f = 1.445\hbox{g}/\hbox{m}^3$ (supersaturation $U_0=0.3$). The inlet velocity is set to $\bm{u}_y= 0.24 \hbox{m}/\hbox{s}$. The evolution of the crystal, stream-lines and supersaturation fields at different time-steps are illustrated in Fig.~\ref{v}.\\
As observed in these figures, the natural anisotropy in supersaturation around the crystal is further accentuated by the formation of two re-circulation zones. Due to the presence of these flow structures, the growth rate on the top half of the crystal is slowed down and brought into the regular hexagonal habit regime. The lower half, however, is subject to larger concentration gradients (further amplified by the incoming convective flux) and therefore exhibits primary branching instabilities. As the system progresses further in time, the bottom-facing branches distinguish themselves from the side-branches, as we observe secondary instability effects on them, along with a much faster growth rate on the main branches. These observations are in clear agreement with expectations from fluid dynamics. Going into the details of the crystal structure and supersaturation fields, we can also see that secondary branching instabilities are not present at the bases of the down-facing primary branches. That is explained by the fact that they are in a flow stagnation zone (due to the built-up pressure in this closed area), where the supersaturation is almost fully depleted. All of these effects are in qualitative agreement with experimental observations reported in the literature \cite{medvedev2005lattice}.
\section{Conclusions and perspectives}
In this study, we presented a \gls{lbm} model for the simulation of snowflakes. Throughout the many generic test-cases it has been shown that this model closely matches results from other solvers for the system of macroscopic \gls{pde}s needed for this system. It was further shown that the proposed formulation is able to capture the different crystal habits in the plate growth regime. The final crystal habits were in good agreement with experimental data and the Nakaya diagram predictions~\cite{nakaya1951formation}. To go one step further, the model was also used to look at the possible effects of forced convection on the growth dynamics and resulting asymmetrical shapes.\\
Given the promising results obtained using this model, future work will focus on the extension of this study to the long-prismatic growth regime to cover the entire spectrum of habits exhibited by snowflakes. Effects from local variations in temperature will also be added to the model to have a better image of the mechanisms behind the growth of snowflakes~\cite{hosseini2019lattice}.
\section{Acknowledgement}
Q.T. would like to acknowledge the financial support by the EU-program ERDF (European Regional Development Fund) within the Research Center for Dynamic Systems (CDS).
S.A.H. would like to acknowledge the financial support of the Deutsche Forschungsgemeinschaft (DFG, German Research Foundation) in TRR 287 (Project-ID 422037413), and thank K. G. Libbrecht for allowing the authors to use images from his ice crystal growth experiments.
\bibliography{main}
\end{document}